\documentclass[12pt]{iopart}

\begin{document}

\title{Mathematical Structure of Discrete Space-time}

\author{An-Wei Zhang}

\address{Department of Modern
Physics, University of Science and Technology of China, Hefei,
Anhui 230026, China}
\ead{zhangzaw@mail.ustc.edu.cn}
\begin{abstract}
In this letter we briefly investigate the mathematical
structure of space-time in the framework of discretization.
It is shown that the discreteness of space-time may result
in a new mechanical system which differ from the usual quantum
mechanics (QM).

\end{abstract}


\section{Introduction}
\label{intro}
Due to the fact that the combined effects of general
relativity and QM does not permit measurement of spatial
or time intervals smaller than the Plank scales~\cite{mead64},
it is generally assumed that space-time should be discrete at such scales.
Actually the concept of a discrete space-time has long
been introduced into physics, as a means of avoiding
the ``infinities'' in theory. Later, it was formulated
by Snyder~\cite{snyder}, Yang~\cite{yang} and Schild~\cite{schild} in the forties.
However, the attempt to construct a basic theory
based on discrete space-time is forced to face many
problems, among which the most extreme one for
our purpose is that the basic laws of physics which
are not easy to be rewritten, such as special relativity
and QM, are established on the basis of
continuous space-time. Therefore, the studies on the discrete
space-time are almost at a standstill for a long time.
Recently, a lot of work has been done, mainly focusing on
some possible properties resulting from the difference
of space-time. And up to now, an appropriate and satisfactory dynamical
system has not been established. An accessible summary
discussion of the development of this concept could be
found in Ref. \cite{gibbs}, in which the major difficulty
of reconstructing QM on discrete space-time, starting from the Hamiltonian formulation
or the path integral formulation, was discussed briefly.

On the other hand, the approach from mathematical angles,
such as taking the space-time as a causal set~\cite{bombelli} and
the measurement on discrete set of points~\cite{brightwell},
which do not provide us with new physics, have some important
results about the properties of space-time, such as the notion
of partial order, which is a basic concept in $C^\star$-algebra,
is introduced. The algebraic approach to the discrete space-time,
such as~\cite{keyl}, attempts to analyze the structure of space-time
with $C^\star$-algebra, unfortunately, a proper dynamical theory
on discrete space-time is not established consequently.

It is necessary for us to present loop quantum gravity, which
has a significant result that certain geometrical quantities,
for instance, area and volume, are represented by operators
that have discrete spectrum~\cite{rovellisomlin}. In~\cite{ashtekar},
Ashtekar and Isham endowed loop transform with a rigorous $C^\star$-algebraic
foundation. It is surprising that distinct theories (the
theory about discrete space-time and the loop quantum gravity
in which time is still taken as a parameter) have similar properties.

In this paper we shall examine that the $C^\dag$-algebra, which is some deformation of
$C^\star$-algebra, can be used to describe the discrete space-time.
We consider an attempt to construct a dynamical equation within the framework of
$C^\dag$-algebra, and point out the differences from the usual Heisenberg equation.
\section{$C^\dag$-algebra}

The concept of discrete space-time can be expressed as
the space and time should be represented by Hermitian
operators that have discrete spectrum at Plank scales.
It can be realized by
\begin{eqnarray}
  \hat{x}|n_x\rangle &=& n\ell_{PL}|n_x\rangle ,\nonumber\\
  \hat{t}|n_t\rangle &=& n\tau_{PL}|n_t\rangle,
\end{eqnarray}
where $|n_x\rangle$ and $|n_t\rangle$ are, respectively,
the eigenket of operator $\hat{x}$ and $\hat{t}$.
We should note the differences from the equivalent continuous
case that the eigenvalue of operator $\hat{x}$ is continous
and the quantity $t$ is taken as a parameter.
The Dirac Notation can still be used, even though the
space-time is discrete, by reason of the fact that
ket or bra only denotes an abstract vector in the mathematical
Hilbert space rather than real physical space. For the sake of
convenience we will require all kets or bras, in this paper, to be normalized.

Since the concept we discussed is quite different from the fundamental
concept of space-time, on which modern physics is constructed, it
is not possible for us to take any laws of physics as the starting
point for the pursuit of dynamics. Naturally, we put our attention toward mathematics,
and hope that a math, which should be nonlocal and non-point, can be found.
Fortunately, as we have seen, there exists an algebra which meets our requirements:
$C^\star$-algebra written as $C^\dag$ here, for ``$\dag$'' satisfies
the requirement of involutive algebra
\begin{eqnarray*}
  (\hat{A}^\dag)^\dag &=& \hat{A}, \\
  (\lambda\hat{A})^\dag  &=& \lambda^\ast \hat{A}^\dag, \\
  (\hat{A}+\hat{B})^\dag  &=& \hat{A}^\dag +\hat{B}^\dag , \\
  (\hat{A}\hat{B})^\dag  &=& \hat{B}^\dag \hat{A}^\dag ,
\end{eqnarray*}
where $\hat{A}^\dag$ is the Hermitian adjoint of $\hat{A}$, $\lambda^\ast$ is
the complex conjugate of $\lambda$ and $\hat{A}$, $\hat{B}\in\mathcal{A}$
which is an algebra.

If the involutive algebra $\mathcal{A}$ is complete and satisfies the relation
\begin{eqnarray*}
  \|\hat{a}^\dag\| &=& \|\hat{a}\|, \\
  \|\hat{a}^\dag\hat{a}\| &=& \|\hat{a}\|^2,
\end{eqnarray*}
$\forall\hat{a}\in\mathcal{A}$, then $\mathcal{A}$ is called a $C^\dag$-algebra, here
 $\|\hat{a}\|$ denotes the spectral radius of $\hat{a}$, which can be
 expressed as $\|\hat{a}\|$=sup$\{$$|\lambda|$; $\hat{a}-\lambda\textbf{1}$ is not invertible$\}$.

 In simple words, let $\mathcal{A}$ be a subset of the set of bounded linear operators
 which is closed under addition, multiplication, multiplication by scalars, the Hermitian
 operation and the operator-norm topology, $\mathcal{A}$ is then called
 a $C^\dag$-algebra~\cite{North-dixmier}. It is worth noting that operator, which only denotes
 an operation acting on a ket, is independent of specific
 space-time.

It is obvious that the operator $\hat{x}$ and $\hat{t}$ satisfy the requirement
of $C^\dag$-algebra, and they, together with other bounded linear Hermitian operators
in physics, can be included in a $C^\dag$-algebra $\mathcal{A}$.

\section{Dynamical equation}

In what follows we shall consider the dynamics on discrete space-time.
we start from an important theorem of $C^\dag$-algebra(see, e.g.,~\cite{Jdix,c-al}):
Let $\mathcal{A}$ be a $C^\dag$-algebra, $\mathcal{B}$ the
weak closure of $\mathcal{A}$, and $\delta$ a derivation of $\mathcal{A}$.
There exists a $\hat{T}\in\mathcal{B}$ such that $\|\hat{T}\|\leq\|\delta\|$ and
\begin{equation}
    \delta \hat{F}=[\hat{T},\hat{F}]
\end{equation}
for every $\hat{F}\in\mathcal{A}$. Where
$\delta$ is a linear derivation, satisfying
$\delta(\hat{a}\hat{b})=(\delta\hat{a})\hat{b}+\hat{a}(\delta\hat{b})$
for all $\hat{a}$, $\hat{b}$$\in\mathcal{A}$.

To get a equation that can generate quantum effect, we multiply
the right side of Eq.(2) by a constant $\frac{i}{\hbar}$, then obtain
\begin{equation}
    \delta \hat{F} = \frac{1}{i\hbar}[\hat{F},\hat{T}],
\end{equation}
where $\hat{T}$ is a conserved quantity.

It is instructive to compare Eq.(3) with the Heisenberg equation
in usual QM. In continuous space-time, one of the most important
features of the dynamical equation is that it involves differentiation
with respect to time or coordinate variable such as the Schr$\ddot{o}$dinger
equation. However, in discrete case, time and space are taken as operators
that have discrete spectrum. Many problems arisen from attempting to
extend the dynamical equation in usual QM to the discrete one. The first
important point is that we do not know whether it makes sense that
we obtain the corresponding dynamical equation just by replacing time or coordinate
variable by corresponding operator or the dynamical equation on discrete space-time
should involve differentiation with respect to the operator $\hat{x}$ or $\hat{t}$.
An interesting feature of the Eq.(3) is the fact that it circumvents such problems
and has a continuous analogue. For this reason, we take Eq.(3) as the dynamical equation
on discrete space-time.

Notice that our approach to Eq.(3) is based solely on $C^\dag$-algebra. It makes sense
whether or not $\hat{F}$ has a continuous counterpart. In other words, due to the fact
that the operator $\hat{t}$ can be included in a $C^\dag$-algebra, so $\hat{t}$ satisfies
\begin{equation}
    \delta \hat{t} = \frac{1}{i\hbar}[\hat{t},\hat{T}].
\end{equation}
However, this equation has no continuous counterpart, because time is taken as a
parameter and cannot be written as operator in usual QM.

\section{Conserved quantity}

To make the present paper self-contained, the next natural
step is to show the explicit form of the operator $\hat{T}$.
For convenience, we introduce an operator $\hat{D}$, which
corresponds to the linear derivation $\delta$. Consider the
fact that $[\hat{D},\hat{F}]|\phi\rangle=\delta(\hat{F}|\phi\rangle)
-\hat{F}(\delta|\phi\rangle)=(\delta\hat{F})|\phi\rangle$, since
$|\phi\rangle$ is an arbitrary ket, we must have
\begin{equation}
   [\hat{D},\hat{F}]=\delta\hat{F}.
\end{equation}

Putting (3) and (5) together, we get
$[\hat{D},\hat{F}]=-\frac{1}{i\hbar}[\hat{T},\hat{F}]$.
Taking into account the fact that $\hat{F}$ is an arbitrary
operator, we can obtain $\hat{T}=-i\hbar\hat{D}$
apart from a constant. Therefore, the conserved quantity
$\hat{T}$ corresponds to $-i\hbar\delta$:
\begin{equation}
    \hat{T}\longrightarrow -i\hbar\delta.
\end{equation}
In continuous limit, compare (6) with $\hat{H}\longrightarrow i\hbar\frac{\partial}{\partial t}$
and $\hat{P}\longrightarrow -i\hbar \nabla$, and we will see that
$\hat{T}$ can be written as
\begin{equation}
    \hat{T} = \varepsilon^i \hat{P}_i - \varepsilon^0 \hat{H}
    =-\varepsilon^\mu \hat{P}_\mu,\nonumber
\end{equation}
which has Lorenze invariance. Here $\varepsilon^\mu$ is an infinitesimal parameter
which denotes space-time translation and $\hat{P}_\mu$ is the 4-momentum.

To illustrate another important formula let us apply the corresponding
relationship (6) to the eigenkets of $\hat{T}$:
\begin{eqnarray}
  -i\hbar\delta|n_T\rangle = \hat{T}|n_T\rangle
   = T_n|n_T\rangle,
\end{eqnarray}
where $|n_T\rangle$ and $T_n$ are, respectively, the eigenket and
eigenvalue of $\hat{T}$. This equation is nothing other than
the generalized Schr\"{o}dinger equation and momentum eigenvalue
equation.

\section{Conclusion}

The properties of space-time at the Plank scales have been studied
for many years, and it is conventionally assumed that the established
laws of physics do not remain valid at such scales. In this paper, our
goal was to construct a systematical quantum theory on discrete space-time.

We introduced a nonlocal $C^\dag$-algebra, which is some deformation
of $C^\star$-algebra, and found that any bounded linear Hermitian operator
in physics can be included in such an algebra. Based solely on $C^\dag$-algebra we obtained the dynamical
equation, which can be generalized to situations where the operators have
no continuous counterpart. If the space-time is
discrete at the Plank scale, its mathematical structure should appear just as
the ones we get in this brief letter.

The author thanks H.X. Yang for useful comment. This work is supported
in part by J.H. Gao.


\section*{References}
\begin{thebibliography}{}
\bibitem{mead64} Mead C A 1964 Possible Connection Between Gravitation and Fundamental Length $\emph{Phys. Rev.}$ $\textbf{135}$ B849
\bibitem{snyder} Snyder H S 1947 Quantized Space-Time $\emph{Phys. Rev.}$ $\textbf{71}$ 38 
\bibitem{yang} Yang C N 1947 On Quantized Space-Time $\emph{Phys. Rev.}$ $\textbf{72}$ 874 
\bibitem{schild} Schild A 1948 Discrete Space-Time and Integral Lorentz Transformations $\emph{Phys. Rev.}$ $\textbf{73}$ 414
\bibitem{gibbs} Gibbs P The Small Scale Structure of Space-Time: A
Bibliographical Review arXiv:hep-th/9506171
\bibitem{bombelli} Bombelli L, Lee J, Meyer D and Sorkin  R D 1987 Space-time as a causal set $\emph{Phys. Rev. Lett.}$ $\textbf{59}$
 521
 \bibitem{brightwell} Brightwell G and Gregory R 1991 Structure of random discrete spacetime $\emph{Phys. Rev. Lett.}$ $\textbf{66}$ 260
 \bibitem{keyl} keyl M 1998 How to describe the space-time structure with nets of $C^\star$-algebras $\emph{Int. J. Theor. Phys.}$ $\textbf{37}$ 375
 \bibitem{rovellisomlin} Rovelli C and Smolin L 1995 Discreteness of area and volume in quantum gravity $\emph{Nucl. Phys. B.}$ $\textbf{442}$ 593
 \bibitem{ashtekar} Ashtekar A and Isham C J 1992 Representations of the holonomy algebras of gravity and nonAbelian gauge theories $\emph{Class. Quantum Grav.}$ $\textbf{9}$ 1433
\bibitem{North-dixmier} Dixmier J 1982 $\emph{$C^\star$-algebras}$ (New York: North-Holland)
\bibitem{Jdix} Dixmier J 1981 $\emph{Von Neumann Algebra}$ (New York: North-Holland) p 349
\bibitem{c-al} Pedersen G K 1979 $\emph{$C^\star$-algebras and Their Automorphism Groups}$ (London: Academic) p 327
\end{thebibliography}
\end{document}